We are commenting on an earlier hypothesis of polaron states bound to F centers in alkali halides. These states increasing the effective size of the color centers, they play an active role in concentration phenomena such as the observed quenching of F center luminescence. Our record shows only one related study on NaBr and NaI which has also been aimed at checking the hypothesis. Further studies of concentration quenching in other alkali halide hosts would eventually throw more light on the problem.

1. Rationale

Some time ago we proposed the concept of a polaron bound to F center [1] in order to explain certain occurrences involving the color centers, in particular the concentration dependence of the F center emission [2], as well as the related concentration dependences of the ionization and deexcitation rates following the optical agitation of the trapped electron [3]. The F-bound polaron concept might also be related to the nature of the chemical bond that holds the extra electron in an F' center. Indeed, with several new F' binding models considered recently, including negative-U [4] and molecular [5] ones, another working suggestion might appear feasible for NaBr and NaI which regards the species as the result of occupying the F-bound polaron states by an additional electron.

Prior to our proposal, 'ghost' polaron states have been reported to occur under conditions of a high excited F center (F*) density provided by a pulsed laser excitation [1,6]. Generally the F-bound polaron ground and excited states are more loosely bound than the corresponding F center states [7]. The polaron potential, self-consistent with the associated lattice distortion, has been represented by an effective truncated Coulomb potential cut off at short range, the so-called polaron radius [8]. In one instance (KI), the ghost polaron absorption has occurred peaking at 0.27 eV while the F center peaks at 1.875 eV.

Casual attempts have subsequently been made for an experimental check up of the F-bound polaron concept by measuring the recovery of the color center absorption following a picosecond pulsed laser bleaching [9]. Two time constants have been resolved, one related to processes "internal" for the relaxing F center and another one characteristic of the "external" ionization of the excited F center. No evidence was reported of any additional process to interfere with the temperature dependence of the internal time constant other than the expected nonradiative deexcitation. Moreover, although the measured F center lifetime has been found to disagree with the lifetime assumed earlier as regards the polaron theory, the authors admit that the disaccord does not have any decisive impact on the that theory.

There are a few comments to be made on the proposed interpretation of the experimental data in [9,10] in view of our recent analyses [11-13] of the nonradiative deexcitation of F centers in NaBr and NaI. They read as follows:

With the coupled vibrational frequency of $\hbar\omega$ = 13 meV and zero-point reaction heat $|Q|$ = 1.3 eV ($Q = Q_2 - Q_1 < 0$ is the difference of the electron binding energies in the excited 2s'-like state ($Q_1$) and the ground 1s-like state ($Q_2$)), there is a vast amount of phonons ($|Q| / \hbar\omega_{vib}$ ~ 100) to be given away to the thermal bath during relaxation. This huge number of phonons reduces the rate of the nonradiative deexcitation process to values beyond measurement. For instance, with a reaction heat of Q ~ $-100$ $\hbar\omega_{vib}$ (exothermic process), the configurational-tunneling probability

$$W_{conf}(E_n) \propto \exp(2Q/\hbar\omega_{vib})$$

is vanishingly small (see equation (13) of ref. [11]). Note also that large $|Q| / \hbar\omega_{vib}$ are inherent to most F centers in alkali halides.

The theoretical nonradiative rate (equation (7) of Ref. [9]) is one of classical nonadiabatic overbarrier transitions across a high crossover barrier at elevated temperatures ($k_BT \gg \hbar\omega_{vib}$) [14]. At lower temperatures ($k_BT \ll \hbar\omega_{vib}$), an apparent temperature along the effective reaction coordinate with frequency $\omega_{eff} \sim \omega_{vib}$ is defined by way of $k_BT^* \sim \frac{1}{2} \hbar\omega_{eff} / \tanh(\hbar\omega_{eff}/2k_BT)$. This procedure leads to a finite zero-point nonradiative rate, as does the reaction rate theory discussed in Refs. [11] & [13]. However, the reaction rate has proved applicable to solid state objects as well [15,16] including the F centers, since it stems directly from the electron-lattice Hamiltonian, while Jortner's theory has originally been aimed at molecules. We are not aware if the 'effective temperature concept' can be derived rigorously from the F center Hamiltonian with a linear electron-vibrational mode coupling and due statistics [17].

The obtained zero-point rate (equation (9) of Ref. [9]) might apply to isothermal processes, since the zero-point reaction heat Q does not enter therein. Moreover, the exponent contains the double barrier $2E_A$ rather than the lattice relaxation energy $E_R \sim 4E_A$ as if to compensate for omitting Q from the zero-point rate ($|Q| = 2E_A$ would do in a particular case). Ultimately, applying molecular arguments to a solid state problem may raise questions.

The foregoing questions and perhaps other ones too stimulated the present comments. In any event, it may be high time to revisit an old problem to see just how it stands today.

2. The F-bound polaron potential and eigenstates

As stated above, what the polaron effectively sees reminds of a radial Coulomb potential cut off at short range defining a polaron radius $r_0$ [8]. The polaron radius has occasionally been estimated independently by the requirement that the electrostatic energy at $r = r_0$ matches the thermal energy of the quasi-particle giving $r_0 \sim e^2/\varepsilon k_BT$. The radial eigenstates are Jacobi functions $j_l(Kr)$ inside the cavity at $r \leq r_0$ followed by hydrogen-like functions $R_{nl}(\alpha r)$ outside the cavity at $r \geq r_0$ [1,7]. The electron-vibrational mode coupling is effected through modulating the polaron radius by the mode coordinate q at a fixed Coulomb tail potential. Under the latter condition the cutoff potential is modulated too in concert with q. The result is a linear electron-$A_{1g}$-mode coupling constant of the form

$$G_{P\psi\psi} = -\omega_{LO}(M_{LO}v_a/4\pi\varepsilon_p e^2)^{1/2} \langle\psi|(\partial/\partial q)[e^2/\varepsilon(r_0+q)]|_{q=0}|\psi\rangle$$

$$= -\omega_{LO}(M_{LO}v_a/4\pi\varepsilon_p e^2)^{1/2} \langle\psi|[e^2/\varepsilon r_0^2]|\psi\rangle \qquad (1)$$

where $|\psi\rangle$ is an electronic state. The factor to the matrix element arises from the $A_{1g}$ lattice mode coupling [7]. Here $v_a$ is the unit cell volume, LO stands for 'longitudinal-optic', $\varepsilon$ and $\varepsilon_p$ are the optic dielectric constant and the polaron constant, respectively.

The semicontinuum approach to the F center is similar to the polaron approach but there is a potential jump at the cutoff radius to account for the Madelung potential $V_M$, followed again by the Coulomb tail [7]. This generates a linear constant in local breathing mode – coupling:

$$G_{F\psi\phi} = \langle\psi|(\partial/\partial q)[V_M r_0/(r_0+q)]|_{q=0}|\psi\rangle = \langle\psi|[V_M/r_0]|\psi\rangle \qquad (2)$$

Both semicontinuum approaches are phenomenological and may be considered inadequate for not providing a deeper insight. Nevertheless, they preserve the physical clarity at some essential points though at the sacrifice of sophisticated mathematics.

The F-bound polaron concept has first been implied while dealing with luminous data in NaBr and NaI [1]. The bound polaron state centered at the anion vacancy enhances the radial size of the F center electron cloud thus making the interaction between neighboring centers more efficient. In a way, the F-bound polaron state is assumed smeared uniformly around the vacancy. Examples for rotational smearing of impurities around isotropic F centers are provided by off-center impurities e.g. at $F_A(Li^+)$ centers in KCl [18].

### 3. Hamiltonian

We introduce a local F center Hamiltonian designed to account for the F-bound polaron field, as follows

$$H = \Sigma_n E_n a_n^\dagger a_n + \tfrac{1}{2} Kq^2 + Gq \Sigma_n a_n^\dagger a_n + \ldots \qquad (3)$$

incorporating electronic, vibrational, and electron-vibrational coupling terms, respectively. Here q, K, and G are the coupled coordinate, stiffness, and coupling constant, respectively, $E_n$ are the electron eigenenergies. The vibrational coordinate q is nonquantized being regarded as a c-number. There are more lattice terms under the dots in addition to those shown above which stay aside by not coupling directly to the electron problem though playing an essential role in taking away the excess energy during relaxation [11], such as

$$\ldots = \tfrac{1}{2} \Sigma_i K_i q_i^2 + \tfrac{1}{2} \Sigma_{ij} K_{ij} q_i q_j + \ldots$$

A semiclassical approach to the lattice problem adopted herein presumes that the coupled coordinates q observe a quantum-mechanical behaviour by exhibiting the zero-point motion. Accordingly we minimize H in q and insert the extremal coordinate thus eliminating q to get

$$q_{ext} = -(G/K) \Sigma_n a_n^\dagger a_n$$

$$H_{ext} = \Sigma_n E_n a_n^\dagger a_n + E_{CE} (\Sigma_n a_n^\dagger a_n)^2 - 2E_{CE} \Sigma_n a_n^\dagger a_n, \qquad (4)$$

the reactive part of the Hamiltonian, where $E_{CE} = \frac{1}{2} G^2/K$ is the coupling energy, $E_n$ is the electron energy (Jacobi-spherical inside and hydrogen-like outside cavity) and $n_n = a_n^\dagger a_n$ are the number of particles operators in the site representation. For a single electron occupation $(\Sigma_n n_n)^2 = \Sigma_n n_n$ and we get

$$H_{ext} = \Sigma_n E_n a_n^\dagger a_n - E_{CE} \Sigma_n a_n^\dagger a_n \equiv \Sigma_n (E_n - E_{CE}) a_n^\dagger a_n \qquad (5)$$

The chief effect of electron-vibrational mode coupling is seen to be the renormalization of the electron energies through lowering by the amount of the coupling energy. This reduces the electron band $\Delta E_n = E_n^\psi - E_n^\phi$ to a squeezed F-bound polaron band, as shown elsewhere [19].

The diabatic potentials (viz. configurational coordinate diagrams) for the vibronic problem are obtained in the adiabatic approximation following the averaging prescription $\langle \Sigma_n a_n^\dagger a_n \rangle = 1$:

$$V_\psi(q) = \frac{1}{2} K_\psi q^2 + G_\psi q + E_\psi, \qquad (6)$$

where $K_\psi = M\omega^2$, $G_\psi$ is the coupling constant, and $E_\psi$ is the electronic eigenvalue in state $\psi$. Examples for diabatic potentials in 1s- and 2p- like F center states and a 2p- like F-bound polaron state are shown in Figure 1 using calculations in Reference [7].

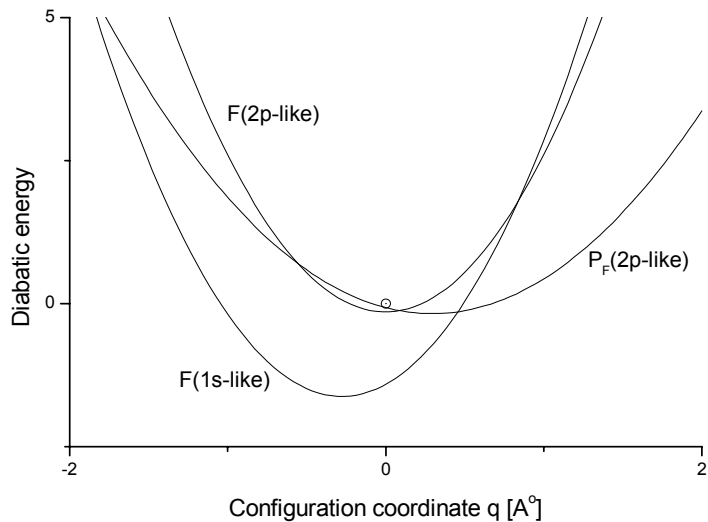

Figure 1:
Configurational coordinate diagram showing diabatic parabolae of 1s- and 2p- like F center states in local breathing-mode coupling, as well as the diabatic parabola of F-bound polaron 2p-like state in $A_{1g}$ lattice-mode coupling. The small circle at the center marks the position of the origin for better viewing. The diagram shows that the F→$P_F$ transfer may be exothermic and proceed with no thermal agitation at the lowest temperatures.

4. Concentration quenching of F center luminescence

Perhaps one of the most important implications of the F-bound polaron concept has been the theoretical view on the observed concentration quenching of the F center luminescence, the ionization quantum efficiency and the F-F' conversion yield [2]. The model assumes that the excited state F* electron cloud detaches spontaneously from the anion vacancy by way of tunnelling to a polaron state $P_F$ bound to a neighboring F center. An $\alpha$-F' pair results which deexcites nonradiatively to a ground-state F-F pair. The F*-$P_F$ transfer occurs as an exothermic phonon-coupled quasi-chemical reaction exhibiting a non-vanishing zero-point rate. The latter rate has no classical analogue and manifests the quantum mechanical nature of the low-temperature process. We also note that the concentration-dependent entity is the cross-over adiabatic energy splitting given by the electronic states' coupling matrix element which is very sensitive to the separation between the interacting centers:

$$V_{FP}(F) = \int V_{FP}(R) \, 4\pi R^2 F \exp(-4\pi/3 \times R^3 F) dR \tag{7}$$

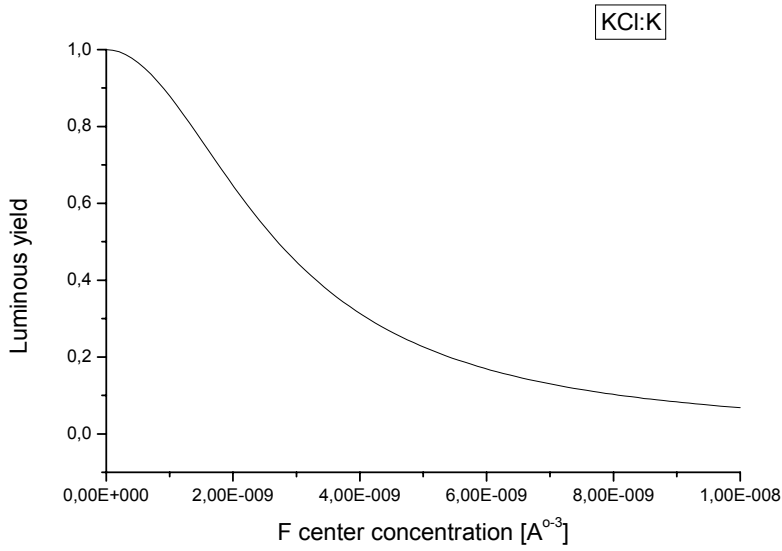

Figure 2:
The concentration quenching of F center luminescence (idealized), as calculated using the on-going equations (7) through (9"). For simplicity the configurational tunnelling is considered to have an occurrence probability of unity.

Accordingly, the luminescent quantum yield is calculated by the polaron theory using

$$\eta_L(F,T) = 1/(1 + k_{FP}(F,T)\tau_F) \tag{8}$$

where $k_{FP}(F,T)$ is the F center – bound polaron transfer rate (to be analyzed next), $\tau_F$ is the F* center lifetime. Performing some of the integration in (7) we get

$$V_{FP}(F) = V_{FP}(0)\{ \int_0^\infty dR \, \exp(-4\pi/3 \times R^3 F) \exp(-\alpha R) - 1\} \tag{9}$$

where $V_{FP}(R) = V_{FP}(0) \exp(-\alpha R)$ stems from the overlap of two ground state hydrogen-like wave functions spaced at R. Series expansion in the first exponential would help integrating a bit further. Using steepest descent, the integral in (9) is approximated to give:

$$V_{FP}(F) = V_{FP}(0)\{ \exp[-(4\pi/3)(1/\alpha^3)F] - 1\} \tag{9'}$$

This matrix element is to be inserted in

$$\gamma(\varepsilon_n, F) = (V_{FP}(F)^2/2\hbar\omega)\varepsilon_R|\varepsilon_C - \varepsilon_n|^{-1/2} = \gamma(\varepsilon_n)\{\exp[-(4\pi/3)(1/\alpha^3)F] - 1\}^2 \qquad (9")$$

and e.g. $W_{el}(\varepsilon_n, F) \sim 4\pi\gamma(\varepsilon_n, F)$ so as to calculate the electronic component probability of a non-adiabatic process [11]. Results of numerical calculations using equations (7) through (9") are presented in Figure 2.

The model [2,20] has been found to agree fairly well with the experimental data on F center luminescence quenching in KCl crystals [21]. The agreement indicates that most F' centers would form by way of intercenter tunneling, to account for their low temperature occurrence. In addition, there may also be a few F' centers formed through F* ionization to conduction band states followed by trapping at F centers, to account for the F' formation in dilute colored crystals. The tunneling picture would give a new insight into the electronic processes in defect crystals.

Further improvements of the theory may involve using quantum-mechanical rather than quasi-classical tunnelling transition probabilities, as well as carrying out calculations on F centers in eight more alkali halides for which a complete set of entry data are available. For the time being, however, we will not go into any more details which can be found in the original publications [2,20].

## 5. Highlights of reaction rate theory

The reaction rate approach (RRA) has been the cornerstone of the F-bound polaron concept. It provides phonon-coupled electronic averages to be used for deriving tunneling probabilities. RRA is based on the occurrence-probability approach to the partial horizontal rates at given vibronic energy levels $\varepsilon_n$ [15]. To avoid misinterpretation, we now repeat some of its basic conclusions. The rate obtains by summing up the partial rates weighed by the thermal occupation factors, as follows:

$$k_{if}(T) = \nu (Z_\omega/Z) \Sigma_n W_n(\varepsilon_n)\exp(-\varepsilon_n/k_BT) \qquad (10)$$

$$= \nu \, 2\sinh(\hbar\omega/2k_BT) \Sigma_{n=0}^\infty W_{\text{conf }n}(\varepsilon_n)W_{\text{el }n}(\varepsilon_n)\exp(-\varepsilon_n/k_BT) \quad [\varepsilon_n = (n+\tfrac{1}{2})\hbar\omega] \qquad (11)$$

$$= \nu \, [1-\exp(-\hbar\omega/k_BT)]\{W_{\text{conf }0}(\varepsilon_0)W_{\text{el }0}(\varepsilon_0) + \Sigma_{n>0}W_{\text{conf }n}(\varepsilon_n)W_{\text{el }n}(\varepsilon_n)\exp(-n\hbar\omega/k_BT)\} \quad (12)$$

The first line is most general as it applies to the quantum mechanical motion along the reactive coordinate of frequency $\nu$ ($\omega=2\pi\nu$), $Z_\omega$ and $Z$ are the partition function of the reactive modes and the total partition function, respectively. To obtain the second line, two addition assumptions are made: (i) harmonic motion along the reactive coordinate, and (ii) Condon's approximation factorizing the overall transition probability $W_n(\varepsilon_n)$ into electronic $W_{eln}(\varepsilon_n)$ and configurational $W_{confn}(\varepsilon_n)$ components. Finally, (iii) is equivalent to (ii) though it manifests clearly the general prediction of a finite zero-point transition rate at n=0:

$$k_{if}(0) = \nu \, W_{\text{conf }0}(\varepsilon_0)W_{\text{el }0}(\varepsilon_0). \qquad (13)$$

The component probabilities are easily derived quasiclassically, though quantum-mechanical expressions are available too [14]. The former are preferred in calculations for the lack of any better alternative. The latter hold good at sub- or over- barrier energy levels that are essentially far from the barrier top. The component probabilities are substantially different for weak [11] and strong coupling [12] configurational coordinate diagrams, as explained therein. For the sake of brevity, we now reproduce the relevant zero-point rates only.

For a strong coupling, when the crossover point is between the equilibrium positions of the pair of diabatic parabolae, we get [13]

$$k_{if}(0)^s = \nu \pi \{[F_{nm}(\xi_0,\xi_C)]^2/2^{n+m}n!m!\} \exp(-Q^2/\varepsilon_R\hbar\omega)\exp(-\varepsilon_R/\hbar\omega) \, 2\pi\gamma^{(2\gamma-1)}\exp(-2\gamma)/[\Gamma(\gamma)]^2 \quad (14)$$

assuming the zero-point transition to start from a subbarrier level. Hereafter $Q = (n-m)\hbar\omega$ is the zero-point reaction heat, $Q < 0$ for exothermic, $Q > 0$ for endothermic, $Q = 0$ for isothermic reaction. $\gamma \equiv \gamma(\varepsilon_n) = (V_{eg}^2/2\hbar\omega)[\varepsilon_R|\varepsilon_n-\varepsilon_C|]^{-1/2}$ is Landau-Zenner's parameter, $F_{nm}$ is a quadratic form of Hermite polynomials [12], $\varepsilon_R$ is the lattice reorganization energy and $\varepsilon_C$ is the crossover energy, $V_{eg}$ is the electron-coupling matrix element (crossover energy splitting).

For a weak coupling, when the crossover point is outside the frame of the pair of equilibrium positions, we have [11]

$$k_{if}(0)^w = \nu \, [(F_{mn}^s/F_{nn}^w)^2(2^n n!)/(2^m m!)]\exp(2Q/\hbar\omega) \, 2\pi\gamma(1-2\gamma)\exp(2\gamma)/\Gamma(1-\gamma) \quad (15)$$

where $F_{nm}^s$ and $F_{nn}^w$ are quadratic forms of Hermite polynomials [11]. The remaining symbols are explained above. Zero-point transitions starting from subbarrier levels are again assumed.

We see that at large $|Q|/\hbar\omega \gg 1$ the zero-point rates are exponentially small, perhaps less so in the strong-coupling case.

## 6. Conclusion

The F-bound polaron has been regarded herein as a set of electronic eigenstates coupled to a lattice vibrational mode(s). We have chosen the $A_{1g}$ mode for the purpose, as done by many, though it is appropriate for describing the compact s-like states mostly. As the F center concentration increases, the individual levels of the electronic eigenstates split to form free-polaron narrow bands. There the distinction between bound- and free- polarons is levelled off.

The F-bound polaron concept has been subjected to, what it looks like, a constructive criticism by a well-known experimental group. Apart from there being no direct evidence for any second activated process at F* in addition to the nonradiative deexcitation, the authors have found a numerical disagreement between a lifetime chosen by the theory and the lifetime observed experimentally, which they themselves regard as nonessential for approving or disapproving the concept. We agree and point out that the problem posed therein, though minor as it is, should stimulate more detailed research. It is also believed that the best way to study the F-bound polaron further is the concentration quenching of luminescence in other F center systems and applying the theory to cover the observed dependences. Unfortunately no novel data are available to us to have appeared during the 16 or so year period since the last published concentration-quenching paper.